\documentclass[11pt]{article}
\usepackage[margin=1in]{geometry}
\usepackage{graphicx} 
\usepackage[utf8]{inputenc}
\usepackage[font=small,labelfont=bf]{caption}
\usepackage{graphicx}
\usepackage{float}
\usepackage{subcaption}
\usepackage[numbers,sort&compress,elide]{natbib}
\usepackage{bm}
\usepackage{ulem}
\usepackage{hyperref}
\usepackage{authblk}  
\usepackage{ amssymb }
\usepackage{cleveref}
\usepackage{bbold}
\usepackage{hyperref}

\crefname{figure}{Fig.}{Figs.}
\Crefname{figure}{Figure}{Figures}

\crefname{equation}{eq.}{eqs.}
\Crefname{equation}{Eq.}{Eqs.}

\hypersetup{colorlinks=true,citecolor=blue, linkcolor=blue,urlcolor=blue}
\graphicspath{{figures/}}   

\begin{document}
\title{Strong Confinement of a Nanodiamond in a Needle Paul Trap: Towards Matter-Wave Interferometry with Massive Objects}
\author{Peter Skakunenko}
\author{Daniel Folman}
\author{Yaniv Bar-Haim}
\author{Ron Folman}

\affil{Ben-Gurion University of the Negev, Department of Physics and Ilse Katz Institute for Nanoscale Science and Technology, Be'er Sheva 84105, Israel}
\maketitle

\begin{abstract} 
Quantum mechanics (QM) and General relativity (GR), also known as the theory of gravity, are the two pillars of modern physics. A matter-wave interferometer with a massive particle, can test numerous fundamental ideas, including the spatial superposition principle - a foundational concept in QM - in completely new regimes, as well as the interface between QM and GR, e.g., testing the quantization of gravity. Consequently, there exists an intensive effort to realize such an interferometer. While several paths are being pursued, we focus on utilizing nanodiamonds as our particle, and a spin embedded in the nanodiamond together with Stern-Gerlach forces, to achieve a closed loop in space-time. There is a growing community of groups pursuing this path \cite{Bose_Morley_2025_white_paper}. We are posting this technical note (as part of a series of seven such notes), to highlight our plans and solutions concerning various challenges in this ambitious endeavor, hoping this will support this growing community. In this work, we achieve strong confinement of a levitated particle, which is crucial for angular confinement, precise positioning, and also advantageous for deep cooling. We designed a needle Paul trap with a controllable distance between the electrodes, giving rise to a strong electric gradient. By combining it with an effective charging method - electrospray - we reach a trap frequency of up to 40\,kHz, which is more than twice the state of the art. We believe that the designed trap could become a significant tool in the hands of the community working towards massive matter-wave interferometry. We would be happy to make more details available upon request.
\end{abstract}

\section{Introduction}

\maketitle


Quantum Mechanics (QM) is a pillar of modern physics. It is thus imperative to test it in ever-growing regions of the relevant parameter space. A second pillar is General relativity (GR), and as a unification of the two seems to be eluding continuous theoretical efforts, it is just as imperative to experimentally test the interface of these two pillars by conducting experiments in which foundational concepts of the two theories must work in concert. 

The most advanced demonstrations of massive spatial superpositions have been achieved by Markus Arndt's group, reaching systems composed of approximately 2,000 atoms\,\cite{Fein2019}. This will surely grow by one or two orders of magnitude in the near future. An important question is whether one can find a new technique that would push the state of the art much further in mass and spatial extent of the superposition. Several paths are being pursued \cite{Romero-Isart_2017,Pino_2018,Weiss2021,Neumeier2024,Kialka2022} and we choose to utilize Stern-Gerlach forces.
The Stern-Gerlach interferometer (SGI) has, in the last decade, proven to be an agile tool for atom interferometry \cite{Amit_2019,dobkowski2025observationquantumequivalenceprinciple,Keil2021}. Consequently, we, as well as others, aim to utilize it for interferometry with massive particles, specifically, nanodiamonds (NDs) with a single spin embedded in their center \cite{Scala_2013,Wan_2016,margalit2021_OUR_intro}.
Levitating, trapping, and cooling of massive particles, most probably a prerequisite for interferometry with such particles, has been making significant progress in recent years. Specifically, the field of levitodynamics is a fast-growing field \cite{gonzalez-ballestero2021_4inOpr}. Commonly used particles are silica spheres. As the state of the art spans a wide spectrum of techniques, achievements and applications, instead of referencing numerous works, we take, for the benefit of the reader, the unconventional step of simply mapping some of the principal investigators; these include Markus Aspelmeyer, Lukas Novotny, Peter Barker, Kiyotaka Aikawa, Romain Quidant, Francesco Marin, Hendrik Ulbricht and David Moore. Relevant to this work, a rather new sub-field which is now being developed deals with ND particles, where the significant difference is that a spin with long coherence times may be embedded in the ND. Such a spin, originating from a nitrogen-vacancy (NV) center, could enable the coherent splitting and recombination of the ND by utilizing Stern-Gerlach forces \cite{Wan_2016,Scala_2013,margalit2021_OUR_intro}. This endeavor includes principal investigators such as Tongcang Li, Gavin Morley, Gabriel Hétet, Tracy Northup, Brian D’Urso, Andrew Geraci, Jason Twamley and Gurudev Dutt.

We aim to start with an ND of $10^7$ atoms and extremely short interferometer durations. Closing a loop in space-time in a very short time is enabled by the strong magnetic gradients induced by the current-carrying wires of the atom chip \cite{Keil2016}. Such an interferometer will already enable to test the existing understanding concerning environmental decoherence (e.g., from blackbody radiation), and internal decoherence \cite{Henkel2024}, never tested on such a large object in a spatial superposition. As we slowly evolve to higher masses and longer durations (larger splitting), the ND SGI will enable the community to probe not only the superposition principle in completely new regimes, but in addition, it will enable to test specific aspects of exotic ideas such as the Continuous spontaneous localization hypothesis \cite{Adler_2021,Gasbarri2021}. As the masses are increased, the ND SGI will be able to test hypotheses related to gravity, such as modifications to gravity at short ranges (also known as the fifth force), as one of the SGI paths may be brought in a controlled manner extremely close to a massive object \cite{Geraci2010,Geraci2015,Bobowski_2024,Panda2024}. Once SGI technology allows for even larger masses ($10^{11}$ atoms), we could test the Diósi–Penrose collapse hypothesis \cite{Penrose2014,Penrose2018,Howl_2019,Tomaz_2024,Bassi_2013} and gravity self-interaction \cite{hatifi2023revealingselfgravitysterngerlachhumptydumpty,Großardt_2021,Aguiar_2024} (e.g., the Schrödinger-Newton equation). Here starts the regime of active masses, whereby not only the gravitation of Earth needs to be taken into account. Furthermore, it is claimed that placing two such SGIs in parallel will allow probing the quantum nature of gravity \cite{Bose2017_quantum_gravity_witness,Marletto_2017}. This will be enabled by ND SGI, as with $10^{11}$ atoms the gravitational interaction could be the strongest \cite{van_de_Kamp_2020,Schut_2023,Schut_2024}. 

Let us emphasize that, although high accelerations may be obtained with multiple spins, we consider only an ND with a single spin as numerous spins will result in multiple trajectories and will smear the interferometer signal. We also note that working with a ND with less than $10^7$ atoms is probably not feasible because of two reasons. The first is that NVs that are closer to the surface than 20\,nm lose coherence, and the second is that at sizes smaller than 50\,nm, the relative fabrication errors become large, and a high-precision ND source becomes beyond reach. 

Here, we present the technical details of our work on one of the building blocks of such an ND SGI, specifically, a Paul trap that enables strong confinement of a levitated ND. This technical note is part of a series of seven technical notes put on the arXiv towards the end of August 2025, including a wide range of required building blocks for the ND SGI \cite{Feldman_Paul_trap_ND, Muretova_ND_theory, Givon_ND_fabrication, Benjaminov_UHV_ND_loading, Liran_ND_neutralization, Levi_Quantum_control_NV}.

\section{Motivation for strong confinement}

For massive matter-wave interferometry, exceptional control of a particle's degrees of freedom is required. Recently, cooling down to the quantum regime was achieved with silica particles in optical tweezers \cite{delic2020aspelmeyer_8inR, magrini2021aspelmeyer_9inR,tebbenjohanns2021novotny_10inR}. In case of diamonds, however, there are issues due to absorption in optical tweezers in high vacuum \cite{Neukirch2015, Hoang2016, Rahman2016, Frangeskou_2018}. Thus, several groups focus their efforts on developing absorption-free Paul traps; in these traps, single-NV manipulation \cite{Conangla2018}, cooling \cite{Conangla2018,jin2024chip}, fast rotation \cite{jin2024chip}, and spin-cooling \cite{delord2020hetet_d} of levitated diamonds in vacuum have been demonstrated. However, deep cooling has not yet been achieved in Paul traps, and as far as we know, the published state of the art stands at about 1\,K\,\cite{jin2024chip} (the achieved temperature was 1.2\,K for an ND with 264\,nm radius). We hypothesize that one of the reasons for this rather high temperature limit is the relatively low trap frequencies of typical Paul traps (up to 15\,kHz \cite{Conangla2018,dania2021_5inO_33in36inR,bonvin2024hybrid}) in comparison with optical tweezers (hundreds of kHz \cite{delic2020aspelmeyer_8inR, magrini2021aspelmeyer_9inR,tebbenjohanns2021novotny_10inR,Aranas2018_thesis}). Traps with higher frequencies enable achieving a higher cooling rate in parametric-feedback cooling schemes \cite{Muretova_ND_theory}, so that the lowest achievable temperature is lower, assuming the limiting factor is heating due to gas collisions. Moreover, the higher the trap frequency, the lower the heating rate due to 1/f noise, inevitable in electronics \cite{Anders_2023_one_over_f}. 

Cooling is important in order to increase the coherence length of the particle. As explained in the so-called Humpty-Dumpty effect (see \cite{margalit2021_OUR_intro} and references therein), in order to achieve a reasonable interference visibility, the coherence length must be larger than the experimental uncertainties in recombining the wavepackets, thus closing the loop in space-time. Our simulations show that for the initial (low mass, short duration) SGI, milli-Kelvin center-of-mass (CoM) temperatures suffice (for rotation, cooling to hundreds of libration phonons is good enough, and this may be relaxed with gyroscopic stabilization, see our work \cite{japha2023_OUR_intro, Muretova_ND_theory}).

In this paper, we demonstrate a method for reaching strong confinement (i.e., high trap frequencies) of a nanoparticle by using a needle Paul trap \cite{Delord_2017, Deslauriers_2006_PRL_Scaling_and} with a controllable distance between the needles, giving rise to high electric gradients.

Let us note that eventually also the Paul trap may be put on the atom chip \cite{jin2024chip}, but for now, the two elements may be placed adjacent to each other, where again the atom chip is required to support the ultra-thin high-current-density current-carrying wire creating the extremely high magnetic gradients which are required for the SG forces. 

Increasing CoM confinement also means increasing torques applied to a levitated particle by the trap fields, as will be shown in the following. The angular confinement due to the Paul trap field torques was already demonstrated with microdiamonds of 15\,$\rm\mu$m size \cite{perdriat2024hetet}. However, angular confinement of nanoparticles (preferred for interferometry because of their smaller mass and moment of inertia) requires much stronger electric gradients together with a high charge-to-mass ratio of the trapped particle, which could be realized with our setup. 
Once an ND with an NV center is angularly confined, it would solve the problem of varying energy levels due to rotation, and precise coherent quantum control can be performed on the embedded NV centers. This is useful not only for massive matter-wave interferometers \cite{Wan_2016, Scala_2013,margalit2021_OUR_intro, japha2023_OUR_intro,Zhou_2024,rusconi2022_my}, but also for other applications such as sensing \cite{jin2024review}. Specific to the ND SGI, rotation also hinders the interference visibility as shown in \cite{japha2023_OUR_intro}. Strong angular confinement is also expected to assist in rotation cooling, as cooling rate of parametric feedback .

 Finally, the stronger confinement is important for massive interferometry for one more reason.  The variance of the particle's position for a given temperature $T$, particle mass $m$ and trap frequency $\omega_{trap}$ is $\langle x^2 \rangle = \frac{k_B T}{m\omega_{trap}^2} $, and consequently, a higher trap frequency makes the particle more localized ("point source") which directly decreases the phase noise of a matter-wave interferometer \cite{dobkowski2025observationquantumequivalenceprinciple,Amit_2019}. Precise positioning (on the order of laser wavelength) is also crucial in cavity-cooling schemes, proven to be effective for ground-state cooling of levitated nanoparticles \cite{delic2020aspelmeyer_8inR, pontin2023barker_30inR, Dania2025}.

\section{Theory}
The potential of a Paul trap can be approximated with the following equation:
\begin{equation}\label{eq:V(r)}
V(x,y,z) = \frac{\eta V_0 \cos(\Omega_{rf} t)}{d^2}[2z^2-(1-\epsilon)y^2-(1+\epsilon)x^2],
\end{equation}
where $V_0$ and $\Omega_{rf}$ are the amplitude and frequency applied to the electrodes, $d$ is the characteristic size of the trap, which is the distance between the needles in case of the needle trap, $\eta$ is the voltage efficiency factor, which defines the reduction of the created field in comparison with the "ideal Paul trap", consisting of infinite hyperbolic electrodes, and $\epsilon$ is the asymmetry factor, quantifying the asymmetry of electrodes in radial dimension. 

A charged particle can be stably trapped if the following condition is fulfilled:

\begin{equation}\label{eq:q_z}
q_z \equiv \frac{8\eta V_0 }{d^2 \Omega_{rf}^2}\frac{Q}{m} < 0.9,
\end{equation}
where $q_z$ is the Mathieu stability parameter along $z$ axis, $Q$ is the total charge of the particle, and $m$ is its mass.
The measure of a particle's confinement in a trap is the trap frequency which is given by
\begin{equation}\label{eq:omega_z}
\omega_z \simeq \frac{q_z \Omega_{rf}}{2\sqrt{2}} =\frac{2 \sqrt{2}\eta V_0 }{d^2 \Omega_{rf}}\frac{Q}{m}
\end{equation}
under the pseudopotential approximation ($q_z \ll 1$). 
The analysis in \cite{perdriat2024hetet} shows that the angular degrees of freedom are described with similar equations \cite{myNote_change_of_axes}:
\begin{equation}\label{eq:q_alfa and omega_alfa}
\omega_{\alpha} \simeq \frac{q_{\alpha}\Omega_{rf}}{2\sqrt{2}}=\frac{2\eta V_0}{3 d^2 \Omega_{rf}} \frac{(3+\epsilon)(Q_3-Q_2)}{I_1},
\end{equation}
where $\omega_\alpha$ and $q_\alpha$ are the libration frequency and libration stability parameter, respectively, $Q_i$ are the eigenvalues of the quadrupole tensor $\hat{Q}=\int \mathrm{d}^3 \mathbf{r} \rho_0(\mathbf{r})\left(3 \mathbf{r} \otimes \mathbf{r}-r^2 \mathbb{1}\right)$ of the particle's charge distribution $\rho_0(\mathbf{r})$ and $I_1$ is the largest moment of inertia.
One can see that both $\omega_z$ and $\omega_{\alpha}$ can be increased by adjusting $V_0$ and $\Omega_{rf}$, which are limited by electronics, and by increasing $\frac{Q}{m}$ ratio, which is limited by the charging mechanism (the highest $\frac{Q}{m}$ is reached with electrospray \cite{Conangla2020_Thesis}). One can also increase angular confinement by using highly elongated particles \cite{perdriat2024hetet}. Finally, another possibility is to decrease the distance between the electrodes, which is the main concept of the needle Paul trap that enables reaching $d$ of tens of microns without sacrificing the $\eta$ factor value.

\section{Experiment}
\begin{figure}[ht!]
\centering
\includegraphics[width=\textwidth,trim=0mm 0mm 0mm 0mm,clip]{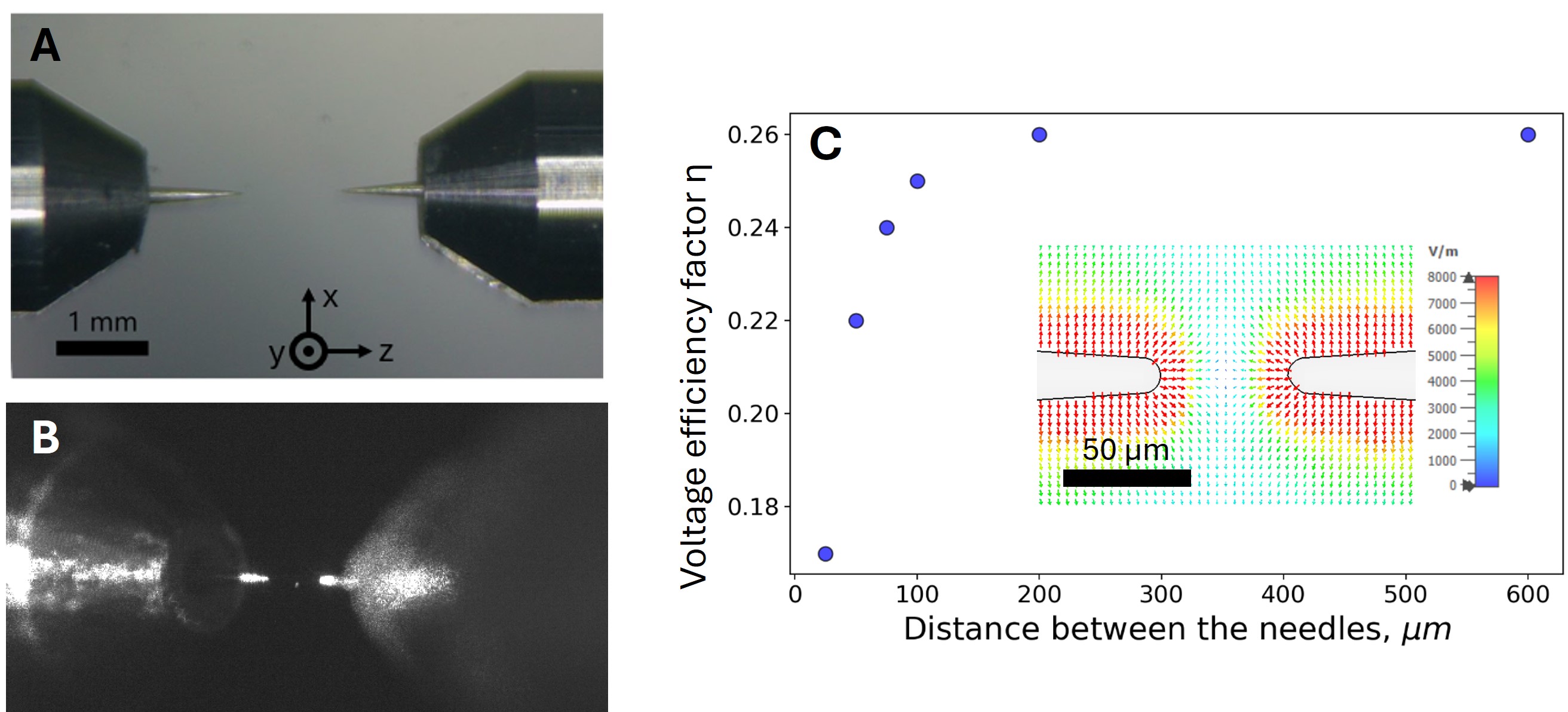}\vspace{-0.2cm}
\caption[Needle trap design and simulations]{Needle trap design and simulations. \textbf{A.} Microscope image of the needle trap. The tungsten steel needles (WG-38.0-10) are fixed inside the ground sleeves and isolated from them by dielectric PEEK tubes. The diameter of the needle tips is approximately $14\,\rm\mu m$. \textbf{B.} ND (Adamas Nanotechnologies, 40\,nm nominal diameter) levitating in the needle Paul trap. The diamond is illuminated with a green laser and observed with a CCD camera at 45 degrees to the needles. The distance between the needles is 730$\,\mu \rm m$. \textbf{C.} Dependence of the voltage efficiency factor, $\eta$, on the distance between the needles, $d$, according to 3D numerical simulations. One can see that, on the one hand, $\eta$ has a reasonable value at large distances and, on the other hand, its decrease at smaller distances (down to $25\,\rm\mu m$) is such that $\frac{\Delta \eta}{\eta}<\frac{\Delta d^2}{d^2}$. Therefore, decreasing $d$ in order to increase the trap frequency [see \Cref{eq:omega_z,eq:q_alfa and omega_alfa}] is still justified. Inset: Electric field distribution in the trap for $d = 50\,\rm\mu m$ at 1\,V DC applied to the needles.
}
\label{fig1}
\end{figure}
The needle Paul trap consists of two sharp needles made of tungsten steel pointing towards each other (\cref{fig1}A), both under the same AC voltage. The needles are fixed in ground sleeves \cite{Deslauriers_2006_PRL_Scaling_and}, which increase the $\eta$ factor by approximately 0.1 and make the voltage well defined. The ground sleeves have notches, designed to create asymmetry to lift the degeneracy of the two radial CoM oscillations.
Around the midpoint between the needles, the field can be approximated by \Cref{eq:V(r)} with $\epsilon \simeq 0.04$ and $\eta$ shown in \cref{fig1}C as a function of $d$.

\begin{figure}[h!]
\centering
\includegraphics[width=1\linewidth,trim=0mm 0mm 0mm 0mm,clip]{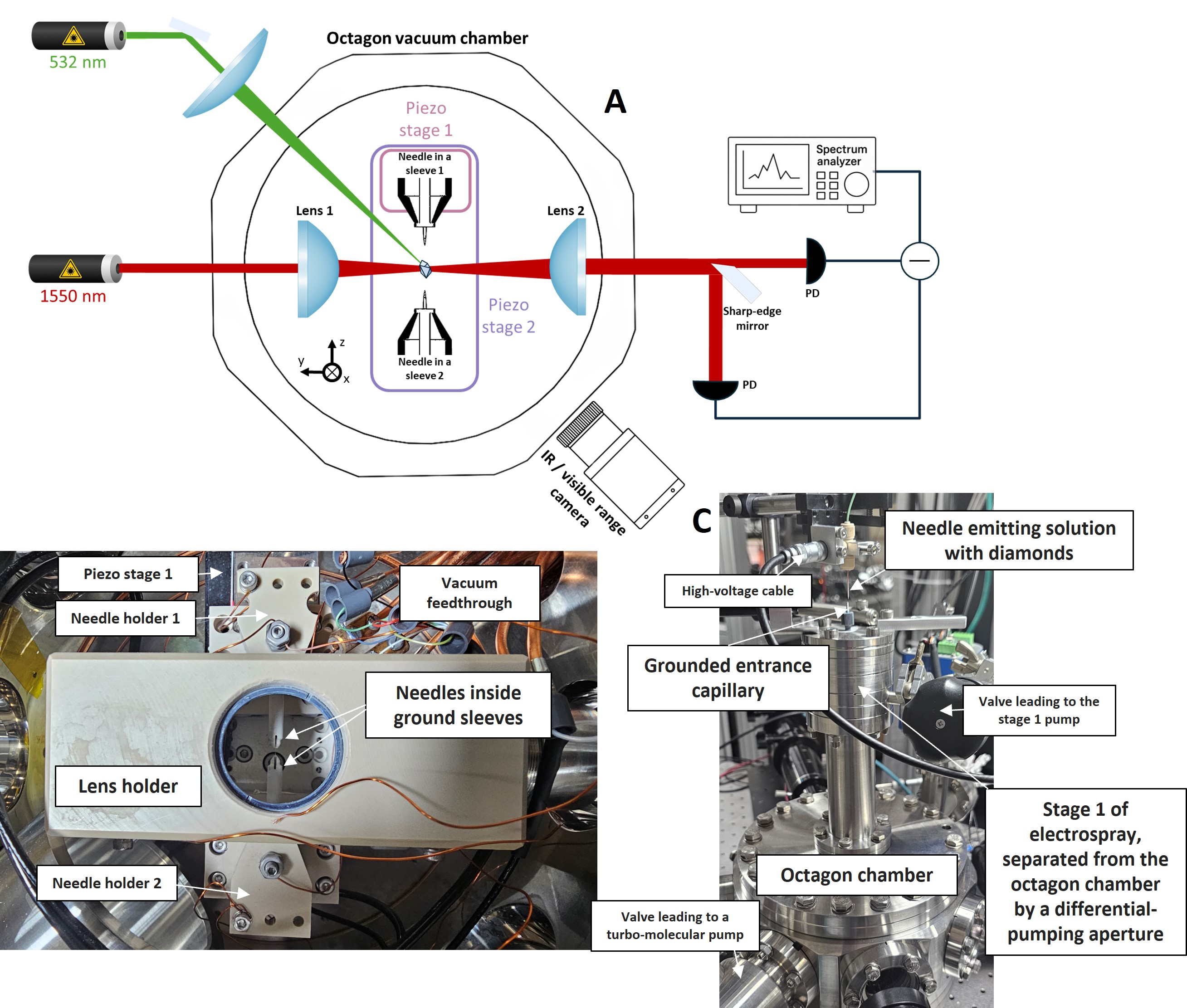}\vspace{-0.2cm}
\caption{The experimental setup. \textbf{A.}
The scheme of the setup. The needles are fixed inside an octagon vacuum chamber on two linear motorized piezo stages (Newport CONEX-AG-LS25-27P), connected to the controllers via vacuum feedthroughs. One of the stages controls the distance between the needles, and another controls the position of both needles. The trap is powered by a signal generator and a voltage amplifier (TREK model 2210). An infrared laser beam ($\lambda = 1550\rm\,nm$) of total power $P\simeq 50\,\rm mW$ is focused into the trap with lens 1 (NA = 0.5). The scattered and unscattered light are collimated by lens 2 (NA = 0.39) and then split by a D-shaped mirror into two inputs of a balanced photodiode. The signal is analyzed with an NI PCI-5922 board serving as a spectrum analyzer. A green laser is used to illuminate the particle. The particle is detected with a visible-range or infrared (IR) camera. \textbf{B.} Photo of the trap inside the octagon chamber. \textbf{C.} Electrospray system. The electrospray system (MolecularSpray UHV4i) is assembled vertically, so that the needle emitting the diamond solution is on top. The high voltage is applied to the solution, so that the charged particles fly from the needle through the grounded entrance capillary (internal diameter 0.25\,mm) inside the vacuum. Entrance capillary leads to the "Stage 1" which is separated from the octagon vacuum chamber with an additional differential-pumping aperture and is connected to a separate rotary pump.
}
\label{setup}
\end{figure}

The trap is located inside a vacuum chamber, and the distance between the needles $d$ is controlled by linear piezo stages (\cref{setup}), with 0.2\,$\rm\mu m$ resolution. The trap can be operated at a wide range of $d$ values from 50\,$\rm\mu m$ to 800\,$\rm\mu m$, which can be seen both from the simulation (\cref{fig1}B) and the experiment. This enables capturing particles at larger $d$, when the trap volume is expanded, and then decreasing $d$ to enhance the confinement. 

One can notice from \Cref{eq:q_z,eq:q_alfa and omega_alfa} that decreasing $d$ by a factor of $p<1$ does not only increases the trap frequencies $\omega_i$ ($i=x,y,z,\alpha,\beta,\gamma$) by the factor of $1/p^2$, but also increases the stability factors $q_i$ by the same factor, which might move the trap out of the stability region. Therefore, the rise of $q_i$ must be compensated by increasing $\Omega_{rf}$ by a factor of $1/p$, so that there is still a gain in trap frequencies of $1/p$. The same logic can be applied to the adjustment of the parameters $Q/m$ and $V_0$. 

The particles are diluted in ethanol and electro-sprayed inside the vacuum chamber through a differential-pumping tube (\cref{setup}C). With such an approach, we manage to trap particles at a pressure down to 3\,Torr, as some amount of damping is needed to decrease the particles' velocity below the escape velocity. After trapping, the pressure can be quickly decreased to below $10^{-4}$ Torr.

The particle's motion is detected with a homodyne forward-scattering technique with split-detection (\cref{setup}). An infrared laser ($\lambda=1550\,\rm nm$) is focused on the particle with a lens (lens 1) and both scattered and unscattered light are collimated with another lens (lens 2), split in half with a D-shaped mirror and sent to the inputs of a balanced photodiode. The signal is analyzed with a spectrum analyzer (see an example of a particle's motion spectrum on \cref{Trap_freq_vs_distance}A).

\section{Results and discussion}

\begin{figure}[H]
\centering
\includegraphics[width=\textwidth,trim=0mm 0mm 0mm 0mm,clip]{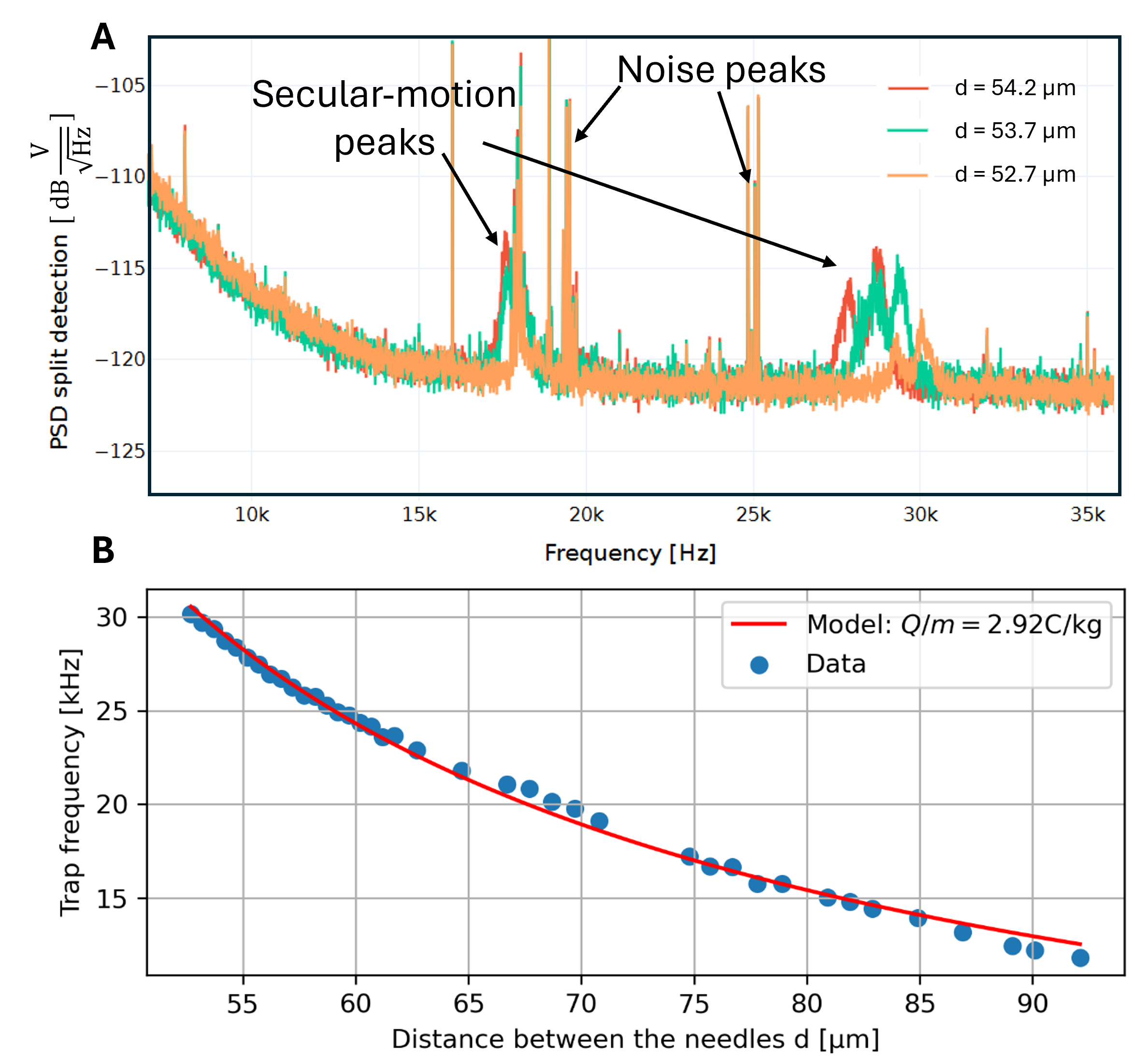}\vspace{-0.2cm}
\caption{Particle's motion spectrum at different distances between the needles. \textbf{A.} Power spectral density (PSD) of the split-detection signal of a diamond nanoparticle at three different distances between the needles $d$. The particle (Adamas Nanotechnologies, 40\,nm nominal diameter) was trapped at a fixed RF voltage and frequency $V_0 = 326\,\rm V_{pp},$ $\Omega_{rf}=2\pi \times 114\,\rm kHz$ and at a fixed pressure of $0.2\rm\,Torr$. The distance between the needles was scanned using the piezo stage 1. The secular-motion peaks are clearly visible; all the other sharper peaks correspond to noise independent of the trap's parameters. \textbf{B.} Maximum trap frequency as a function of the distance between the needles. The model takes into account higher-order corrections beyond the pseudopotential approximation according to \cite{Lindvall_2022}, because $q_z$ reaches a value of 0.75 at the smallest distances. The dependence $\eta(d)$ is also taken into account. The model has two free parameters: the charge-to-mass ratio of the particle estimated by the fit as $Q/m = 2.92\,\rm C/kg$, and DC-shift of the applied voltage, estimated to be $U_0=9\,\rm V$ (the latter coincides with the independently measured value). See \hyperref[Appendix]{Appendix} for the model's details. We observed jumps of the trap frequency between two values separated by approximately 1\,kHz, likely corresponding to two stable equilibrium positions, but further study is needed. The graph shows the highest-frequency mode.}
\label{Trap_freq_vs_distance}
\end{figure}

\Cref{Trap_freq_vs_distance}B shows the trap frequency of a diamond particle (Adamas Nanotechnologies, 40\,nm nominal diameter) as a function of the distance between the needles $d$. As can be seen, we reached a trap frequency of more than 30\,kHz (another particle reached 40\,kHz), which is, to the best of our knowledge, at least twice as large as the state-of-the-art value for nanoparticles levitated in Paul traps \cite{Conangla2018}. The diameter of the particle, estimated from the PSD peak width \cite{magrini2021aspelmeyer_9inR}, equals $63.6 \pm 3.8$\,nm, which is within the manufacturer's uncertainty. These values allow us to estimate the particle's minimal uncertainty in position to be $ \sigma_z \approx \sqrt\frac{k_B T}{m\omega_{trap}^2} \approx\rm370\,nm $ along the axis of the strongest confinement.

During the measurement, the pressure was fixed at 0.2\,Torr, because at lower pressure the probability of losing the particle increased. The probable reason for this is the heating due to laser scattering. We find that when the laser power is set to a lower value of $\approx$ 100\,$\mu$W, the particle can stay trapped at the pressure of $10^{-3}$\,Torr and, probably, lower. However, in our setup, SNR becomes insufficient at this laser power. Other detection schemes, such as back-scattering detection, heterodyne detection, or PBS detection, as well as higher NA optics, can be applied in order to increase the SNR.

\section{Outlook}
The trap frequency can be increased even more by further decreasing the distance between the needles. In our case, the limiting factor is the diffraction of the detection beam off the needles, preventing effective detection. This can be solved by decreasing the laser spot size and using sharper needles. Heating due to laser absorption is an issue that needs to be addressed by high-purity diamonds. Another possible limiting factor is the heating rate caused by anomalous heating, which can be compensated with cryogenic needles \cite{Deslauriers_2006_PRL_Scaling_and}. 

If deep cooling is one's main aim, we expect to observe an optimal distance between the needles at which the benefit from the high trap frequency is not yet canceled by the increase in anomalous heating rate (which scales as $\sim1/d^4$ \cite{Boldin2018,Deslauriers_2006_PRL_Scaling_and}).

As was already mentioned, the angular confinement can be increased by using highly non-spherical particles. This will be achieved by utilizing an ND source made of precisely etched NDs. In our fabrication work, we have made significant progress in this respect \cite{Givon_ND_fabrication}.

The $Q/m$ ratio can be increased in situ by UV-light ionization of a positively charged particle or by irradiating a negatively charged particle with electrons, by focusing the UV light on the adjacent electrodes (see our work with UV \cite{Liran_ND_neutralization}).

This work is part of our effort to demonstrate massive quantum superpositions using NDs with NV centers (see our NV work \cite{Levi_Quantum_control_NV}) and SG forces for closing a loop in space-time (ND SGI). Next, we will integrate this trap into our other setups, which have already achieved cooling and deep vacuum \cite{Feldman_Paul_trap_ND}, and bring the atom chip to the vicinity of ND to generate strong magnetic gradients. In parallel, we will also study planar (on-chip) high-frequency Paul traps for integration in more advanced generations of the ND SGI.

\section*{Acknowledgments}
We thank the BGU Atom-Chip Group support team, especially Menachem Givon, Zina Binstock, Dmitrii Kapusta and Michael Averbukh for their support in building and maintaining the experiment. We also thank Or Dobkowski and Omer Feldman from our group, and Marko Toroš from the University of Ljubljana, for useful discussions. Funding: This work has been supported by the "Table-top experiments for fundamental physics" program, sponsored by the Gordon and Betty Moore Foundation, Simons Foundation, Alfred P. Sloan Foundation, and John Templeton Foundation.

\section*{Appendix. Equations of motion in a Paul trap and corrections beyond the pseudopotential approximation}\label{Appendix}
The expressions below follow \cite{Lindvall_2022}. \Cref{eq:V(r)} from the main text neglects the possible DC-component of the Paul trap field. The full expression gives:
\begin{equation}\label{eq:V_full(r)}
V(x,y,z) = \frac{\eta_{DC}U_0+\eta V_0 \cos(\Omega_{rf} t)}{d^2}[2z^2-(1-\epsilon)y^2-(1+\epsilon)x^2] + V_{ext}(x,y,z),
\end{equation}
where $U_0$ is DC-shift of the trap voltage, $\eta_{DC}$ is corresponding voltage efficiency factor and $V_{ext}(x,y,z)$ is the potential of external stray fields. Note that in the following, $V_{ext}$ is assumed to be small enough, so that the two radial motions can be described by two decoupled one-dimensional Mathieu equations. If it is not the case and both RF and DC components of the field are highly asymmetric along different axes, then the radial motion cannot be decoupled into two one-dimensional motions (see \cite{Lindvall_2022} for details).

For decoupled axes, the equations of motion are written in the form of the Mathieu equations:
\begin{equation}\label{eq:Mathieu}
\frac{d^2 x_i}{d \tau^2}+\left(a_i-2 q_i \cos 2 \tau\right) x_i=0,
\end{equation}
where $\tau = \Omega_{rf}t/2$, $q_z = \frac{8\eta V_0 }{d^2 \Omega_{rf}^2}\frac{Q}{m}$, $q_{x,y}=-(1\pm\epsilon)q_z/2$, $a_z=-\frac{16\eta_{DC} U_0 }{d^2 \Omega_{rf}^2}\frac{Q}{m} + a^{ext}_{z}$, $a_{x,y}=-(1\pm\epsilon)(a_z-a^{ext}_{z})/2 + a^{ext}_{x,y}$. 

The stable solution of \cref{eq:Mathieu} is:
\begin{equation}\label{eq:Mathieu_solution}
x_i(t)=C \sum_{n=-\infty}^{\infty} c_{2 n} \cos \left[\left(2 n+\beta_i\right) \tau+\phi_i\right].
\end{equation}
Parameters $\beta_i$ define secular frequencies $\omega_i=\beta_i\Omega_{rf}/2$ and can be found numerically, $C$ and $\phi_i$ depend on initial conditions, and $c_n$ are given by recursion relations \cite{major2005charged_book}. To lowest order in the parameters $q_i$ and $a_i$, the motion becomes a multiplication of two motions: micromotion at frequency $\Omega_{rf}$ and macromotion at secular frequencies $\omega_i$:
\begin{equation}\label{App_eq:motion_sec_approx}
x_i(t) \approx C_1 \cos \left(\omega_i t+\phi_i\right)\left(1-\frac{q_i}{2} \cos \Omega t\right),
\end{equation}
with the secular frequencies given by:
\begin{equation}\label{App_eq:secular_frequencies}
\omega_i=\frac{\Omega_{rf}}{2} \beta_i \approx \frac{\Omega_{rf}}{2} \sqrt{a_i+q_i^2 / 2}.
\end{equation}
The above approximation is called the pseudopotential approximation, as the micromotion has a much smaller amplitude than the macromotion, so that the potential can be considered as a harmonic trap with frequency $\omega_i$.

However, when the condition $q_i\ll1$ is not fulfilled, the correction to the pseudopotential approximation is needed. Expanding expression for $\beta_i$ to order $a_i^1$, $q_i^6$ yields:
\begin{equation}\label{App_eq:beta_expanded}
 \beta_i^2 \approx a_i+\left(\frac{1}{2}+\frac{a_i}{2}\right) q_i^2+\left(\frac{25}{128}+\frac{273}{512} a_i\right) q_i^4 +\left(\frac{317}{2304}+\frac{59525}{82944} a_i\right) q_i^6 .
\end{equation}
This approximation gives an error of below 2\% for our parameters.
\Cref{App_eq:beta_expanded} and \cref{App_eq:secular_frequencies} were used to fit the experimental data in \cref{Trap_freq_vs_distance} with two fitting parameters: the charge-to-mass ratio of the particle and DC offset, returning the values $Q/m=2.92\,\rm C/kg$ and $U_0=9\rm\,V$ (the latter coincides with the separately measured value).


\begin{thebibliography}{10}
\newcommand{\enquote}[1]{``#1''}

\bibitem{Bose_Morley_2025_white_paper}
S.~Bose, A.~Mazumdar, R.~Penrose, \textit{et al.}, \href{https://arxiv.org/abs/2509.01586}{\enquote{A spin-based pathway to testing the quantum nature of gravity,} }preprint arXiv:2509.01586 (2025).

\bibitem{Fein2019}
Y.~Y. Fein, P.~Geyer, P.~Zwick, F.~Kia{\l}ka, S.~Pedalino, M.~Mayor, S.~Gerlich, and M.~Arndt, \href{https://doi.org/10.1038/s41567-019-0663-9}{\enquote{{Quantum superposition of molecules beyond 25 kDa},} }Nature Physics \textbf{15}, 1242--1245 (2019).

\bibitem{Romero-Isart_2017}
O.~Romero-Isart, \href{https://dx.doi.org/10.1088/1367-2630/aa99bf}{\enquote{Coherent inflation for large quantum superpositions of levitated microspheres,} }New Journal of Physics \textbf{19}, 123029 (2017).

\bibitem{Pino_2018}
H.~Pino, J.~Prat-Camps, K.~Sinha, B.~P. Venkatesh, and O.~Romero-Isart, \href{https://dx.doi.org/10.1088/2058-9565/aa9d15}{\enquote{On-chip quantum interference of a superconducting microsphere,} }Quantum Science and Technology \textbf{3}, 025001 (2018).

\bibitem{Weiss2021}
T.~Weiss, M.~Roda-Llordes, E.~Torrontegui, M.~Aspelmeyer, and O.~Romero-Isart, \href{https://link.aps.org/doi/10.1103/PhysRevLett.127.023601}{\enquote{Large quantum delocalization of a levitated nanoparticle using optimal control: Applications for force sensing and entangling via weak forces,} }Physical Review Letters \textbf{127}, 023601 (2021).

\bibitem{Neumeier2024}
L.~Neumeier, M.~A. Ciampini, O.~Romero-Isart, M.~Aspelmeyer, and N.~Kiesel, \href{https://www.pnas.org/doi/abs/10.1073/pnas.2306953121}{\enquote{Fast quantum interference of a nanoparticle via optical potential control,} }Proceedings of the National Academy of Sciences \textbf{121}, e2306953121 (2024).

\bibitem{Kialka2022}
F.~Kiałka, Y.~Y. Fein, S.~Pedalino, S.~Gerlich, and M.~Arndt, \href{https://doi.org/10.1116/5.0080940}{\enquote{A roadmap for universal high-mass matter-wave interferometry,} }AVS Quantum Science \textbf{4}, 020502 (2022).

\bibitem{Amit_2019}
O.~Amit, Y.~Margalit, O.~Dobkowski, Z.~Zhou, Y.~Japha, M.~Zimmermann, M.~A. Efremov, F.~A. Narducci, E.~M. Rasel, W.~P. Schleich, and R.~Folman, \href{https://link.aps.org/doi/10.1103/PhysRevLett.123.083601}{\enquote{${T}^{3}$ stern-gerlach matter-wave interferometer,} }Physical Review Letters \textbf{123}, 083601 (2019).

\bibitem{dobkowski2025observationquantumequivalenceprinciple}
O.~Dobkowski, B.~Trok, P.~Skakunenko, Y.~Japha, D.~Groswasser, M.~Efremov, C.~Marletto, I.~Fuentes, R.~Penrose, V.~Vedral, W.~P. Schleich, and R.~Folman, \href{https://arxiv.org/abs/2502.14535}{\enquote{Observation of the quantum equivalence principle for matter-waves,} }preprint arXiv:2502.14535  (2025).

\bibitem{Keil2021}
M.~Keil, S.~Machluf, Y.~Margalit, Z.~Zhou, O.~Amit, O.~Dobkowski, Y.~Japha, S.~Moukouri, D.~Rohrlich, Z.~Binstock, Y.~Bar-Haim, M.~Givon, D.~Groswasser, Y.~Meir, and R.~Folman, \href{https://link.springer.com/book/10.1007/978-3-030-63963-1}{\enquote{Stern-gerlach interferometry with the atom chip,} }Molecular Beams in Physics and Chemistry: From Otto Stern's Pioneering Exploits to Present-Day Feats pp. 263--301 (2021).

\bibitem{Scala_2013}
M.~Scala, M.~S. Kim, G.~W. Morley, P.~F. Barker, and S.~Bose, \href{https://link.aps.org/doi/10.1103/PhysRevLett.111.180403}{\enquote{Matter-wave interferometry of a levitated thermal nano-oscillator induced and probed by a spin,} }Physical Review Letters \textbf{111}, 180403 (2013).

\bibitem{Wan_2016}
C.~Wan, M.~Scala, G.~W. Morley, A.~A. Rahman, H.~Ulbricht, J.~Bateman, P.~F. Barker, S.~Bose, and M.~S. Kim, \href{https://link.aps.org/doi/10.1103/PhysRevLett.117.143003}{\enquote{Free nano-object ramsey interferometry for large quantum superpositions,} }Physical Review Letters \textbf{117}, 143003 (2016).

\bibitem{margalit2021_OUR_intro}
Y.~Margalit, O.~Dobkowski, Z.~Zhou, O.~Amit, Y.~Japha, S.~Moukouri, D.~Rohrlich, A.~Mazumdar, S.~Bose, C.~Henkel, and R.~Folman, \href{https://www.science.org/doi/full/10.1126/sciadv.abg2879}{\enquote{Realization of a complete stern-gerlach interferometer: Toward a test of quantum gravity,} }Science Advances \textbf{7}, eabg2879 (2021).

\bibitem{gonzalez-ballestero2021_4inOpr}
C.~Gonzalez-Ballestero, M.~Aspelmeyer, L.~Novotny, R.~Quidant, and O.~Romero-Isart, \href{https://www.science.org/doi/abs/10.1126/science.abg3027}{\enquote{Levitodynamics: Levitation and control of microscopic objects in vacuum,} }Science \textbf{374}, eabg3027 (2021).

\bibitem{Keil2016}
M.~Keil, O.~Amit, S.~Zhou, D.~Groswasser, Y.~Japha, and R.~Folman, \href{http://dx.doi.org/10.1080/09500340.2016.1178820}{\enquote{{Fifteen years of cold matter on the atom chip: promise, realizations, and prospects},} }Journal of Modern Optics \textbf{63}, 1840--1885 (2016).

\bibitem{Henkel2024}
C.~Henkel and R.~Folman, \href{https://link.aps.org/doi/10.1103/PhysRevA.110.042221}{\enquote{{Universal limit on spatial quantum superpositions with massive objects due to phonons},} }Physical Review A \textbf{110}, 42221 (2024).

\bibitem{Adler_2021}
S.~L. Adler, A.~Bassi, and M.~Carlesso, \href{https://dx.doi.org/10.1088/1751-8121/abdbc8}{\enquote{The continuous spontaneous localization layering effect from a lattice perspective,} }Journal of Physics A: Mathematical and Theoretical \textbf{54}, 085303 (2021).

\bibitem{Gasbarri2021}
G.~Gasbarri, A.~Belenchia, M.~Carlesso, S.~Donadi, A.~Bassi, R.~Kaltenbaek, M.~Paternostro, and H.~Ulbricht, \href{https://doi.org/10.1038/s42005-021-00656-7}{\enquote{{Testing the foundation of quantum physics in space via Interferometric and non-interferometric experiments with mesoscopic nanoparticles},} }Communications Physics \textbf{4}, 155 (2021).

\bibitem{Geraci2010}
A.~A. Geraci, S.~B. Papp, and J.~Kitching, \href{https://link.aps.org/doi/10.1103/PhysRevLett.105.101101}{\enquote{Short-range force detection using optically cooled levitated microspheres,} }Physical Review Letters \textbf{105}, 101101 (2010).

\bibitem{Geraci2015}
A.~Geraci and H.~Goldman, \href{https://link.aps.org/doi/10.1103/PhysRevD.92.062002}{\enquote{Sensing short range forces with a nanosphere matter-wave interferometer,} }Physical Review D \textbf{92}, 062002 (2015).

\bibitem{Bobowski_2024}
J.~S. Bobowski, H.~Patel, and M.~Faizal, \href{https://dx.doi.org/10.1088/1402-4896/ad3178}{\enquote{Novel setup for detecting short-range anisotropic corrections to gravity,} }Physica Scripta \textbf{99}, 045017 (2024).

\bibitem{Panda2024}
C.~D. Panda, M.~J. Tao, M.~Ceja, J.~Khoury, G.~M. Tino, and H.~M{\"{u}}ller, \href{https://doi.org/10.1038/s41586-024-07561-3}{\enquote{{Measuring gravitational attraction with a lattice atom interferometer},} }Nature \textbf{631}, 515--520 (2024).

\bibitem{Penrose2014}
R.~Penrose, \href{https://doi.org/10.1007/s10701-013-9770-0}{\enquote{On the gravitization of quantum mechanics 1: Quantum state reduction,} }Foundations of Physics \textbf{44}, 557--575 (2014).

\bibitem{Penrose2018}
R.~Penrose and I.~Fuentes, \href{https://www.cambridge.org/core/books/collapse-of-the-wave-function/FD0F1EC77C4949379B63B0643B7AC7D4}{\enquote{Quantum state reduction via gravity, and possible tests using bose–einstein condensates,} }Collapse of the Wave Function: Models, Ontology, Origin, and Implications pp. 187--206 (2018).

\bibitem{Howl_2019}
R.~Howl, R.~Penrose, and I.~Fuentes, \href{https://dx.doi.org/10.1088/1367-2630/ab104a}{\enquote{Exploring the unification of quantum theory and general relativity with a bose–einstein condensate,} }New Journal of Physics \textbf{21}, 043047 (2019).

\bibitem{Tomaz_2024}
A.~A. Tomaz, R.~S. Mattos, and M.~Barbatti, \href{http://dx.doi.org/10.1039/D4CP02364A}{\enquote{Gravitationally-induced wave function collapse time for molecules,} }Physical Chemistry Chemical Physics \textbf{26}, 20785--20798 (2024).

\bibitem{Bassi_2013}
A.~Bassi, K.~Lochan, S.~Satin, T.~P. Singh, and H.~Ulbricht, \href{https://link.aps.org/doi/10.1103/RevModPhys.85.471}{\enquote{Models of wave-function collapse, underlying theories, and experimental tests,} }Reviews of Modern Physics \textbf{85}, 471--527 (2013).

\bibitem{hatifi2023revealingselfgravitysterngerlachhumptydumpty}
M.~Hatifi and T.~Durt, \href{https://arxiv.org/abs/2006.07420}{\enquote{Revealing self-gravity in a stern-gerlach humpty-dumpty experiment,} }preprint arXiv:2006.07420  (2023).

\bibitem{Großardt_2021}
A.~Großardt, \href{https://dx.doi.org/10.1088/1361-6382/ac36a6}{\enquote{Dephasing and inhibition of spin interference from semi-classical self-gravitation,} }Classical and Quantum Gravity \textbf{38}, 245009 (2021).

\bibitem{Aguiar_2024}
G.~H.~S. Aguiar and G.~E.~A. Matsas, \href{https://link.aps.org/doi/10.1103/PhysRevA.109.032223}{\enquote{Probing the schr\"odinger-newton equation in a stern-gerlach-like experiment,} }Physical Review A \textbf{109}, 032223 (2024).

\bibitem{Bose2017_quantum_gravity_witness}
S.~Bose, A.~Mazumdar, G.~W. Morley, H.~Ulbricht, M.~{Toro\ifmmode \checks\else {\v{s}}\fi}, M.~Paternostro, A.~A. Geraci, P.~F. Barker, M.~S. Kim, and G.~Milburn, \href{https://link.aps.org/doi/10.1103/PhysRevLett.119.240401}{\enquote{{Spin Entanglement Witness for Quantum Gravity},} }Physical Review Letters \textbf{119}, 240401 (2017).

\bibitem{Marletto_2017}
C.~Marletto and V.~Vedral, \href{https://link.aps.org/doi/10.1103/PhysRevLett.119.240402}{\enquote{Gravitationally induced entanglement between two massive particles is sufficient evidence of quantum effects in gravity,} }Physical Review Letters \textbf{119}, 240402 (2017).

\bibitem{van_de_Kamp_2020}
T.~W. van~de Kamp, R.~J. Marshman, S.~Bose, and A.~Mazumdar, \href{https://link.aps.org/doi/10.1103/PhysRevA.102.062807}{\enquote{Quantum gravity witness via entanglement of masses: Casimir screening,} }Physical Review A \textbf{102}, 062807 (2020).

\bibitem{Schut_2023}
M.~Schut, A.~Grinin, A.~Dana, S.~Bose, A.~Geraci, and A.~Mazumdar, \href{https://link.aps.org/doi/10.1103/PhysRevResearch.5.043170}{\enquote{Relaxation of experimental parameters in a quantum-gravity-induced entanglement of masses protocol using electromagnetic screening,} }Physical Review Research \textbf{5}, 043170 (2023).

\bibitem{Schut_2024}
M.~Schut, A.~Geraci, S.~Bose, and A.~Mazumdar, \href{https://link.aps.org/doi/10.1103/PhysRevResearch.6.013199}{\enquote{Micrometer-size spatial superpositions for the qgem protocol via screening and trapping,} }Physical Review Research \textbf{6}, 013199 (2024).

\bibitem{Feldman_Paul_trap_ND}
O.~Feldman, B.~B. Shultz, M.~Muretova, O.~Dobkowski, Y.~Japha, D.~Groswasser, and R.~Folman, \href{https://arxiv.org/abs/2508.14687}{\enquote{Trapping and cooling of nanodiamonds in a Paul trap under ultra-high vacuum: Towards matter-wave interferometry with massive objects,} } preprint arXiv:2508.14687 (2025).

\bibitem{Muretova_ND_theory}
M.~Muretova, Y.~Japha, M.~Toro\v{s}, and R.~Folman, \href{https://arxiv.org/abs/2508.13723}{\enquote{Parametric feedback cooling of librations of a nanodiamond in a Paul trap: Towards matter-wave interferometry with massive objects,} } preprint arXiv:2508.13723 (2025).

\bibitem{Givon_ND_fabrication}
M.~Givon, Y.~Bar-Haim, D.~Groswasser, A.~Solodar, N.~Aharon, M.~Belman, A.~Yosefi, E.~Golan, J.~Jopp, and R.~Folman, \href{https://arxiv.org/abs/2508.13662}{\enquote{Fabrication of nano-diamonds with a single {NV} center: Towards matter-wave interferometry with massive objects,} } preprint arXiv:2508.13662 (2025).

\bibitem{Benjaminov_UHV_ND_loading}
R.~Benjaminov, S.~Liran, O.~Dobkowski, Y.~Bar-Haim, M.~Averbukh, and R.~Folman, \href{https://arxiv.org/abs/2508.14722}{\enquote{Design of high-efficiency UHV loading of nanodiamonds into a Paul trap: Towards matter-wave interferometry with massive objects,} } preprint arXiv:2508.14722 (2025).

\bibitem{Liran_ND_neutralization}
S.~Liran, O.~Dobkowski, R.~Benjaminov, P.~Skakunenko, M.~Averbukh, Y.~Bar-Haim, D.~Groswasser, J.~H. Baraban, and R.~Folman, \href{https://arxiv.org/html/2508.15625}{\enquote{Neutralization of levitated charged nanodiamond: Towards matter-wave interferometry with massive objects,} } preprint arXiv:2508.15625 (2025).

\bibitem{Levi_Quantum_control_NV}
N.~Levi, O.~Feldman, Y.~Rosenzweig, D.~Groswasser, A.~Elgarat, M.~Gal-Katziri, and R.~Folman, \href{https://www.arxiv.org/abs/2508.15504}{\enquote{Quantum control of nitrogen-vacancy spin in nanodiamonds: Towards matter-wave interferometry with massive objects,} } preprint arXiv:2508.15504 (2025).

\bibitem{delic2020aspelmeyer_8inR}
U.~Delić, M.~Reisenbauer, K.~Dare, D.~Grass, V.~Vuletić, N.~Kiesel, and M.~Aspelmeyer, \href{https://www.science.org/doi/abs/10.1126/science.aba3993}{\enquote{Cooling of a levitated nanoparticle to the motional quantum ground state,} }Science \textbf{367}, 892--895 (2020).

\bibitem{magrini2021aspelmeyer_9inR}
L.~Magrini, P.~Rosenzweig, C.~Bach, A.~Deutschmann-Olek, S.~G. Hofer, S.~Hong, N.~Kiesel, A.~Kugi, and M.~Aspelmeyer, \href{https://www.nature.com/articles/s41586-021-03602-3#citeas}{\enquote{Real-time optimal quantum control of mechanical motion at room temperature,} }Nature \textbf{595}, 373--377 (2021).

\bibitem{tebbenjohanns2021novotny_10inR}
F.~Tebbenjohanns, M.~L. Mattana, M.~Rossi, M.~Frimmer, and L.~Novotny, \href{https://www.nature.com/articles/s41586-021-03617-w#citeas}{\enquote{Quantum control of a nanoparticle optically levitated in cryogenic free space,} }Nature \textbf{595}, 378–382 (2021).

\bibitem{Neukirch2015}
L.~P. Neukirch, E.~von Haartman, J.~M. Rosenholm, and A.~{Nick Vamivakas}, \href{https://doi.org/10.1038/nphoton.2015.162}{\enquote{{Multi-dimensional single-spin nano-optomechanics with a levitated nanodiamond},} }Nature Photonics \textbf{9}, 653--657 (2015).

\bibitem{Hoang2016}
T.~M. Hoang, J.~Ahn, J.~Bang, and T.~Li, \href{https://doi.org/10.1038/ncomms12250}{\enquote{{Electron spin control of optically levitated nanodiamonds in vacuum},} }Nature Communications \textbf{7}, 12250 (2016).

\bibitem{Rahman2016}
A.~T. M.~A. Rahman, A.~C. Frangeskou, M.~S. Kim, S.~Bose, G.~W. Morley, and P.~F. Barker, \href{https://doi.org/10.1038/srep21633}{\enquote{{Burning and graphitization of optically levitated nanodiamonds in vacuum},} }Scientific Reports \textbf{6}, 21633 (2016).

\bibitem{Frangeskou_2018}
A.~C. Frangeskou, A.~T. M.~A. Rahman, L.~Gines, S.~Mandal, O.~A. Williams, P.~F. Barker, and G.~W. Morley, \href{https://dx.doi.org/10.1088/1367-2630/aab700}{\enquote{Pure nanodiamonds for levitated optomechanics in vacuum,} }New Journal of Physics \textbf{20}, 043016 (2018).

\bibitem{Conangla2018}
G.~P. Conangla, A.~W. Schell, R.~A. Rica, and R.~Quidant, \href{https://doi.org/10.1021/acs.nanolett.8b01414}{\enquote{{Motion Control and Optical Interrogation of a Levitating Single Nitrogen Vacancy in Vacuum},} }Nano Letters \textbf{18}, 3956--3961 (2018).

\bibitem{jin2024chip}
Y.~Jin, K.~Shen, P.~Ju, X.~Gao, C.~Zu, A.~J. Grine, and T.~Li, \href{https://doi.org/10.1038/s41467-024-49175-3}{\enquote{Quantum control and berry phase of electron spins in rotating levitated diamonds in high vacuum,} }Nature Communications \textbf{15}, 5063 (2024).

\bibitem{delord2020hetet_d}
T.~Delord, P.~Huillery, L.~Nicolas, and G.~H{\'e}tet, \href{https://doi.org/10.1038/s41586-020-2133-z}{\enquote{Spin-cooling of the motion of a trapped diamond,} }Nature \textbf{580}, 56--59 (2020).

\bibitem{dania2021_5inO_33in36inR}
L.~Dania, D.~S. Bykov, M.~Knoll, P.~Mestres, and T.~E. Northup, \href{https://link.aps.org/doi/10.1103/PhysRevResearch.3.013018}{\enquote{Optical and electrical feedback cooling of a silica nanoparticle levitated in a paul trap,} }Physical Review Research \textbf{3}, 013018 (2021).

\bibitem{bonvin2024hybrid}
E.~Bonvin, L.~Devaud, M.~Rossi, A.~Militaru, L.~Dania, D.~S. Bykov, M.~Teller, T.~E. Northup, L.~Novotny, and M.~Frimmer, \href{https://doi.org/10.1103/PhysRevResearch.6.043129}{\enquote{Hybrid paul-optical trap with large optical access for levitated optomechanics,} }Physical Review Research \textbf{6}, 043129 (2024).

\bibitem{Aranas2018_thesis}
E.~Aranas, \href{https://api.semanticscholar.org/CorpusID:198448385}{\enquote{Levitated optomechanics with periodically driven fields,} }Ph.D. thesis, UCL (University College London) (2018).

\bibitem{Anders_2023_one_over_f}
J.~Anders, M.~Babaie, J.~C. Bardin, I.~Bashir, G.~Billiot, E.~Blokhina, S.~Bonen, E.~Charbon, J.~Chiaverini, I.~L. Chuang, C.~Degenhardt, D.~Englund, L.~Geck, L.~Le~Guevel, D.~Ham, R.~Han, M.~I. Ibrahim, D.~Krüger, K.~M. Lei, A.~Morel, D.~Nielinger, G.~Pillonnet, J.~M. Sage, F.~Sebastiano, R.~B. Staszewski, J.~Stuart, A.~Vladimirescu, P.~Vliex, and S.~P. Voinigescu, \href{http://dx.doi.org/10.1109/TQE.2023.3290593}{\enquote{Cmos integrated circuits for the quantum information sciences,} }IEEE Transactions on Quantum Engineering \textbf{4}, 1--30 (2023).

\bibitem{japha2023_OUR_intro}
Y.~Japha and R.~Folman, \href{https://doi.org/10.1103/PhysRevLett.130.113602}{\enquote{Quantum uncertainty limit for stern-gerlach interferometry with massive objects,} }Physical Review Letters \textbf{130}, 113602 (2023).

\bibitem{Delord_2017}
T.~Delord, L.~Nicolas, L.~Schwab, and G.~H{\'{e}}tet, \href{https://dx.doi.org/10.1088/1367-2630/aa659c}{\enquote{{Electron spin resonance from NV centers in diamonds levitating in an ion trap},} }New Journal of Physics \textbf{19}, 33031 (2017).

\bibitem{Deslauriers_2006_PRL_Scaling_and}
L.~Deslauriers, S.~Olmschenk, D.~Stick, W.~K. Hensinger, J.~Sterk, and C.~Monroe, \href{https://link.aps.org/doi/10.1103/PhysRevLett.97.103007}{\enquote{Scaling and suppression of anomalous heating in ion traps,} }Physical Review Letters \textbf{97}, 103007 (2006).

\bibitem{perdriat2024hetet}
M.~Perdriat, C.~C. Rusconi, T.~Delord, P.~Huillery, C.~Pellet-Mary, A.~Durand, B.~A. Stickler, and G.~H{\'e}tet, \href{https://doi.org/10.1103/PhysRevLett.133.253602}{\enquote{Rotational locking of charged microparticles in quadrupole ion traps,} }Physical Review Letters \textbf{133}, 253602 (2024).

\bibitem{Zhou_2024}
R.~Zhou, R.~J. Marshman, S.~Bose, and A.~Mazumdar, \href{https://dx.doi.org/10.1088/1402-4896/ad37df}{\enquote{Gravito-diamagnetic forces for mass independent large spatial superpositions,} }Physica Scripta \textbf{99}, 055114 (2024).

\bibitem{rusconi2022_my}
C.~C. Rusconi, M.~Perdriat, G.~H\'etet, O.~Romero-Isart, and B.~A. Stickler, \href{https://link.aps.org/doi/10.1103/PhysRevLett.129.093605}{\enquote{Spin-controlled quantum interference of levitated nanorotors,} }Physical Review Letters \textbf{129}, 093605 (2022).

\bibitem{jin2024review}
Y.~Jin, K.~Shen, P.~Ju, and T.~Li, \href{https://arxiv.org/abs/2407.12496}{\enquote{Towards real-world applications of levitated optomechanics,} }preprint arXiv:2407.12496  (2024).

\bibitem{pontin2023barker_30inR}
A.~Pontin, H.~Fu, M.~Toroš, T.~S. Monteiro, and P.~F. Barker, \href{https://www.nature.com/articles/s41567-023-02006-6}{\enquote{Simultaneous cavity cooling of all six degrees of freedom of a levitated nanoparticle,} }Nature Physics \textbf{19}, 1003–1008 (2023).

\bibitem{Dania2025}
L.~Dania, O.~S. Kremer, J.~Piotrowski, D.~Candoli, J.~Vijayan, O.~Romero-Isart, C.~Gonzalez-Ballestero, L.~Novotny, and M.~Frimmer, \href{https://doi.org/10.1038/s41567-025-02976-9}{\enquote{{High-purity quantum optomechanics at room temperature},} }Nature Physics  (2025).

\bibitem{myNote_change_of_axes}
$\alpha, \beta, \gamma$ are the Euler angles in $zy'z''$ convention. In order to attain this result, the axes should be changed, so that $y$ is the axis of the strongest gradient, $\epsilon>0$. In the body frame (when $\alpha = \beta = \gamma = 0$) the particle shape is chosen such that the inertia momenta satisfy $I_1>I_2>I_3$. The approximation is made around the angular position $\alpha=0,\beta=\pi/2,\gamma=0$.

\bibitem{Conangla2020_Thesis}
G.~Planes~Conangla, \href{https://hdl.handle.net/2117/329754}{\enquote{Levitation and control of particles with internal degrees of freedom,} }Universitat Polit{\`e}cnica de Catalunya, Barcelona (2020).

\bibitem{Lindvall_2022}
T.~Lindvall, K.~J. Hanhijärvi, T.~Fordell, and A.~E. Wallin, \href{http://dx.doi.org/10.1063/5.0106633}{\enquote{High-accuracy determination of paul-trap stability parameters for electric-quadrupole-shift prediction,} }Journal of Applied Physics \textbf{132} (2022).

\bibitem{Boldin2018}
I.~A. Boldin, A.~Kraft, and C.~Wunderlich, \href{https://link.aps.org/doi/10.1103/PhysRevLett.120.023201}{\enquote{Measuring anomalous heating in a planar ion trap with variable ion-surface separation,} }Physical Review Letters \textbf{120}, 023201 (2018).

\bibitem{major2005charged_book}
F.~Major, V.~Gheorghe, and G.~Werth, \href{https://books.google.co.il/books?id=E5NxU-0nf5oC}{\enquote{Charged particle traps: Physics and techniques of charged particle field confinement,} }Springer Berlin Heidelberg  (2005).

\end{thebibliography}
\end{document}